# Monocular passive event-based range-finding of airborne objects using the Scheimpflug principle

Nathan Meraz[a,*], Ronan Taneja[a], Rachel Chan[a], Alisha Whitehead[a], Gabriella Mayrend[a], Megan Birch[a], and Joseph L. Greene[a]
[a]Georgia Tech Research Institute, Atlanta, GA, 30318, USA
*Nathan.Meraz@gtri.gatech.edu

## ABSTRACT

Passive 3D sensing is increasingly critical for early detection and tracking of small aerial vehicles (UAVs), where traditional active ranging can be tactically undesirable. We present SCHeimpflug for Optical Ranging TechnologY (SCHORTY), a single-aperture passive and active ranging architecture that exploits the Scheimpflug principle to encode range along a tilted object space plane by tilting the sensor relative to the imaging optics. SCHORTY requires only a one-time geometric calibration to map pixel coordinates to range and is inherently sensor and waveband agnostic. We implement SCHORTY using both a visible frame-based camera and an event-based camera (EBC) with closely matched pixel sizes for comparable horizontal resolutions and range binning. Controlled flights of an octocopter and a fixed-wing UAV equipped with GPS provide ground truth distances out to 1.1 km. Experimental results show that SCHORTY achieves deterministic range assignment limited primarily by the projected pixel size, which grows squared distance, while avoiding computationally intensive inverse reconstructions common in coded aperture and PSF engineered systems. In the EBC configuration, EBC-SCHORTY inherently suppresses static background and emphasizes motion, improving UAV detectability in cluttered natural scenes and under turbulence and motion blur. Additionally, we observe an asymmetric defocus blur about the object plane that depends on UAV trajectory, suggesting an extra cue for localization and trajectory inference. These results demonstrate SCHORTY as a practical and Size, Weight, and Power (SWaP) efficient passive ranging solution for medium-range UAV observation and motivate future integration with 2.5D/3D PSF engineering and event-based deconvolution to enhance 3D sensing performance.

## 1. INTRODUCTION

Recovering 3D positional information from compressed 2D images serves as a fundamental challenge when developing vision systems that may sufficiently identify and assess incoming targets [1]. This challenge is compounded with modern aerial objects, such as unmanned aerial vehicles (UAVs), which use irregular shapes, short deployment ranges, and swarm-based tactics to confound existing surveillance systems [2]. These effects lead to an existing gap when innovate optical systems that may efficiently detect aerial objects at ranges sufficient for early detection. Active solutions, such as LiDARs, provide high-resolution 3D information by monitoring the return from emitted lasers, but lead to operational vulnerability by compromising the emitter location [3]. To mitigate this limitation, recent work highlights passive modalities, which leverage the physical encoding of range information by select optical architectures without the need for a known reference [4].

In general, passive systems fall into three categories: multi-aperture, sub-aperture, and single-aperture (i.e. monocular) techniques. Multi-aperture techniques, such as parallax or triangulation, synthesize 3D information by measuring angular disparity achieved between physically separated imaging systems with discrete viewpoints [5]. In principle, these methods may flexibly adapt by adjusting the baseline between platforms to achieve a required resolution and scale in aperture to achieve a required range. In practice, encamping and synchronizing multiple platforms leads increased operational cost and necessitate downstream data fusion, which increasing the latency and compute associated with recovering 3D information. Sub-aperture solutions split an aperture into multiple viewpoints to achieve parallax with a single system, yet sacrifice spatial resolution on the sensor plane, leading to an undesirable tradeoff between lateral and axial resolution [6].

Single-aperture solutions prioritize preserving the SWaP and resolution of a vision system, but necessitate additional optics to modify the optical transfer function to permit range encoding. These methods included coded aperture and PSF engineering [7], [8], [9], [10]. Both methods introduce amplitude or phase encoding mechanisms to modify how light propagates through an imaging system to create axially encoded features to computationally recover range. However, the design and optimization of these optical cues is highly ill-poised and necessitate computationally intensive reconstruction algorithms, which exhibit sensitivities to real-world effects such as the presence of noise or confusers [11], [12]. As a result, the field will benefit from solutions that balances the SWaP benefits of single-aperture approaches but provides robust, customizable 3D encoding without sacrificing simplicity or incurring complex reconstruction.

To meet these demands, this work introduces a passive 3D imaging sensing platform based on the Scheimpflug principle [13], titled the SCHeimpflug for Optical Ranging TechnologY (SCHORTY) system. This principle leverages the tilting of a camera plane respective to a planar lens to map pixels to a tilted plane in object space. By controlling the degree of tilt and imaging system parameters (i.e., focal length, pixel size), SCHORTY achieves 100 m range binning at 1 km, which is approximately half the hyperfocal distance. Due to the simple mechanism to enable ranging, this platform only requires a one-time calibration to account for aberrations or assembly errors to properly map pixel coordinate to range [14]. By solely requiring a tilted sensor plane to enable ranging, SCHORTY natively adapts to image sensors across modalities and wavebands to provide a flexible architecture to enable 3D sensing of objects within the field-of-view (FoV). To emphasize how SCHORTY may combine with leading research to promote enhance detection of UAVs, we combine SCHORTY with an event-based camera (EBC) [15] to benefit from the ability to encode motion while suppressing the effects from previously limiting challenges, such as background, turbulence, and motion blur [16], [17], [18]. Our demonstration fills a gap between SWaP efficient single aperture systems and practical passive ranging by providing experimentally validated ranging for UAVs out to 1 km, without relying on ill-posed coded or PSF-engineered reconstructions. In doing so, SCHORTY establishes a scalable camera-neutral approach for 3D encoding.

# 2. METHODS

## 2.1 Design of a Passive Scheimpflug 3D Ranging System

SCHORTY is designed using the Scheimpflug principle to map pixel coordinates along a tilted range axis in 3D space, following the theory and calibrations in [14]. In brief, SCHORTY comprises a lens (Rokinon F/2.0, 135 mm focal length), an arrayed sensor, and a custom 3D printed mount to hold the sensor at a 20-degree angle relative to the lens plane. SCHORTY is deployed using a visible frame-based camera (Basler ACE acA1300-200um) and an asynchronous EBC (Prophesee EVK4). These cameras are selected because they exhibit comparable pixel sizes and horizontal resolutions (Basler: 4.8µm pixel size, 1280 x 1024 resolution. Prophesee: 4.86 µm pixel size, 1280 x 720 resolution) such that each may map to the same ranges with similar range bins when placed in the Scheimpflug geometry.

Under the 20-degree tilt, both systems map from 30 meters to 2 km, which is bounded by the Scheimpflug distance offset and hyperfocal range of the platform, respectively. Due to the extended range, SCHORTY requires objects and scales beyond laboratory conditions to calibrate its range. To calibrate the cameras, SCHORTY was encamped at a known GPS coordinate and imaged a landscape. GPS coordinates were collected at distinguishable landmarks (e.g., trees, buildings, radio towers) and those distances were converted into a range through Haversine's formula for comparison to the ranging predicted by SCHORTY. Sensor models were used to compare the ranges and predict the as-built sensor tilt angle and offset. Due to the native rejection of static sources inhibiting the recording of fixed objects, the EBC SCHORTY initially assumed the same range mapping as the visible camera. After analyzing the time-synced EBC recordings and GPS tracks, the EBC SCHORTY pixel-range calibration was adjusted to fit the observed pixel coordinates to known UAV distances.

## 2.2 Execution of Controlled Drone Flights

Relevant aerial objects (e.g., an industrial octocopter UAV with a 70” wingspan & a commercial fixed-wing drone with a 77” wingspan) were deployed with an active antenna GPS puck to monitor position. The octocopter was automated on predetermined flight trajectories automated using the Ground Control Station (GCS) software while the fixed-wing was manually operated. GCS was used with a GNSS receiver to extract ground truth drone coordinates during operation. Drone coordinates were compared against SCHORTY's encamped GPS coordinate to convert drone position into geographic distance and serve as ground truth for SCHORTY measurements.

# 3. RESULTS

## 3.1 Example Calibration of SCHORTY Against Known Landmarks at Range

To calibrate SCHORTY, the platform (see Figure 1B) was encamped at a known location and its GPS coordinate recorded. Next, landmarks were registered within the FoV for calibration and their GPS coordinates recorded. As shown in Figure 1A, the GPS coordinates estimate range along the shortest distance path between the two points. When placed along the Scheimpflug axis (see Figure 1A Bottom), in-focus features fall creating sharp features within the geometric depth of field at a range corresponding to its pixel coordinate. By comparing the measured range to the predicted range, model agreement may be determined (see Figure 1C).

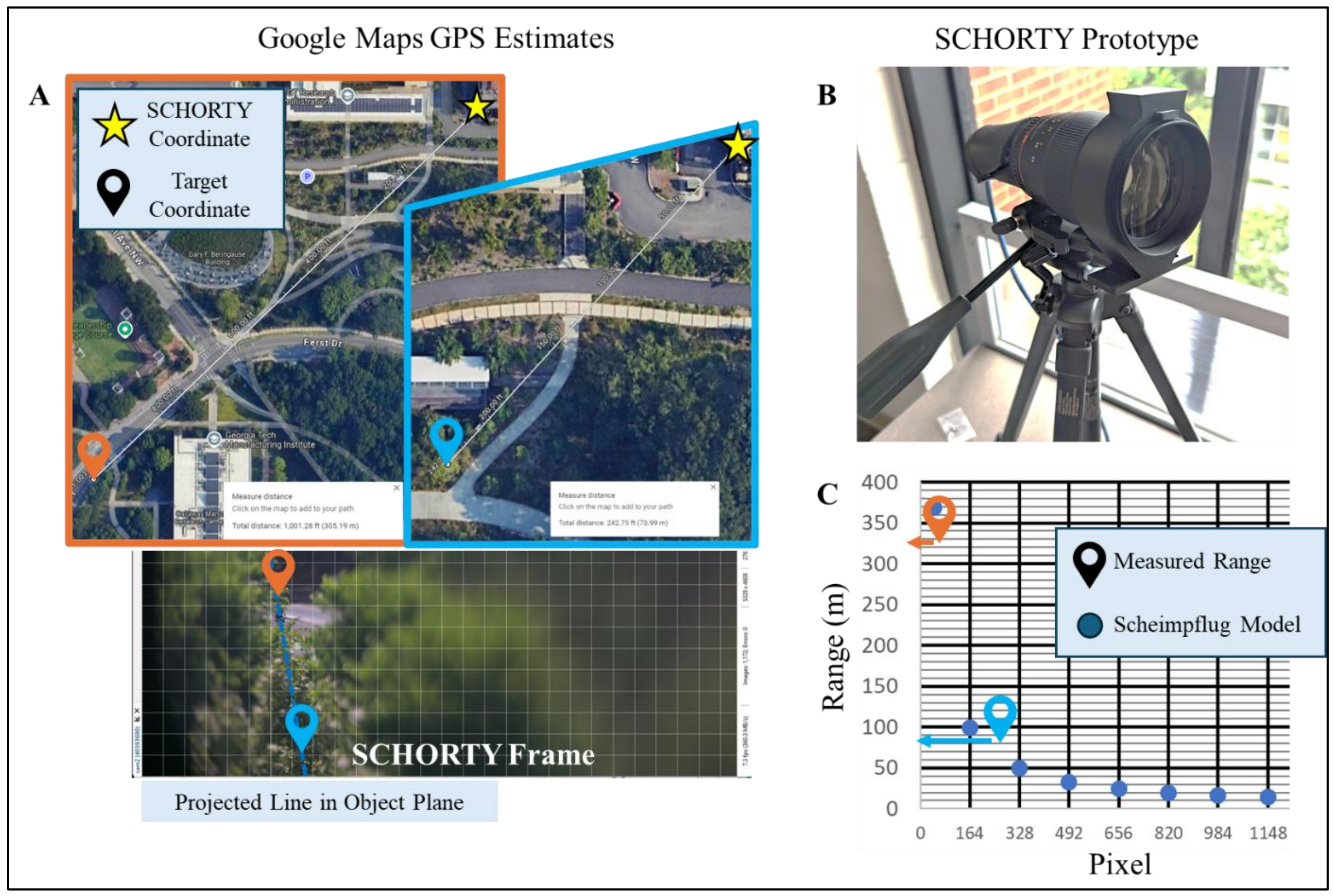

Figure. 1. Calibration of a SCHORTY Rangefinding Platform. (A) SCHORTY's position is compared against registered landmarks to determine range. (B) SCHORTY system used in study. (C) Comparison of measured landmarks to predicted range mapping by Scheimpflug principle.

## 3.2 Measured Comparison of SCHORTY vs GPS for Drone Tracking

To assess SCHORTY performance on a fast-moving target with an extended flight profile, we conducted a series of range finding experiments using fixed wing and octocopter UAVs. The UAVs repeatedly traversed the depth mapped object plane while the SCHORTY systems were positioned a fixed location. Figure 2A and 3A show sequential SCHORTY images and selected EBC-SCHORTY event frames of the different UAVs as they approached and intersected the in-focus object plane. As each UAV's projected position along the object plane changes, its features transition from blurred to sharp with higher contrast. The pixel-range calibration maps provide the range measurement based on the row where best focus is observed. Figure 2C and 3C compare the GPS-derived line of sight distance to three predicted ranges from each system. These measurements agree well with the expected distances which validates the calibration and confirms that range can be assigned deterministically over a 1 km scale from a single frame without inverse construction. These plots include an estimate on the distance-dependent depth of field which represents a typical limitation to camera-based focus assessment.

Experiments were conducted with SCHORTY deployed in three configurations: mounted on a fixed tripod, mounted on a motorized 2-axis gimbal, and operated as a handheld instrument. In each case, the targets were manually tracked to assess operator ability to keep UAVs within the depth mapped plane and to demonstrate that practical ranging and tracking can be achieved without reliance on high-performance dedicated tracking mounts. Figure 2D and 3D illustrate the recovered 3D trajectory by combining the derived ranges with the gimbal's encoded azimuth and elevation angles to compute the UAV's position in a local coordinate frame. The GPS flight path is overlaid onto the three SCHORTY and EBC-SCHORTY track positions with good correspondence.

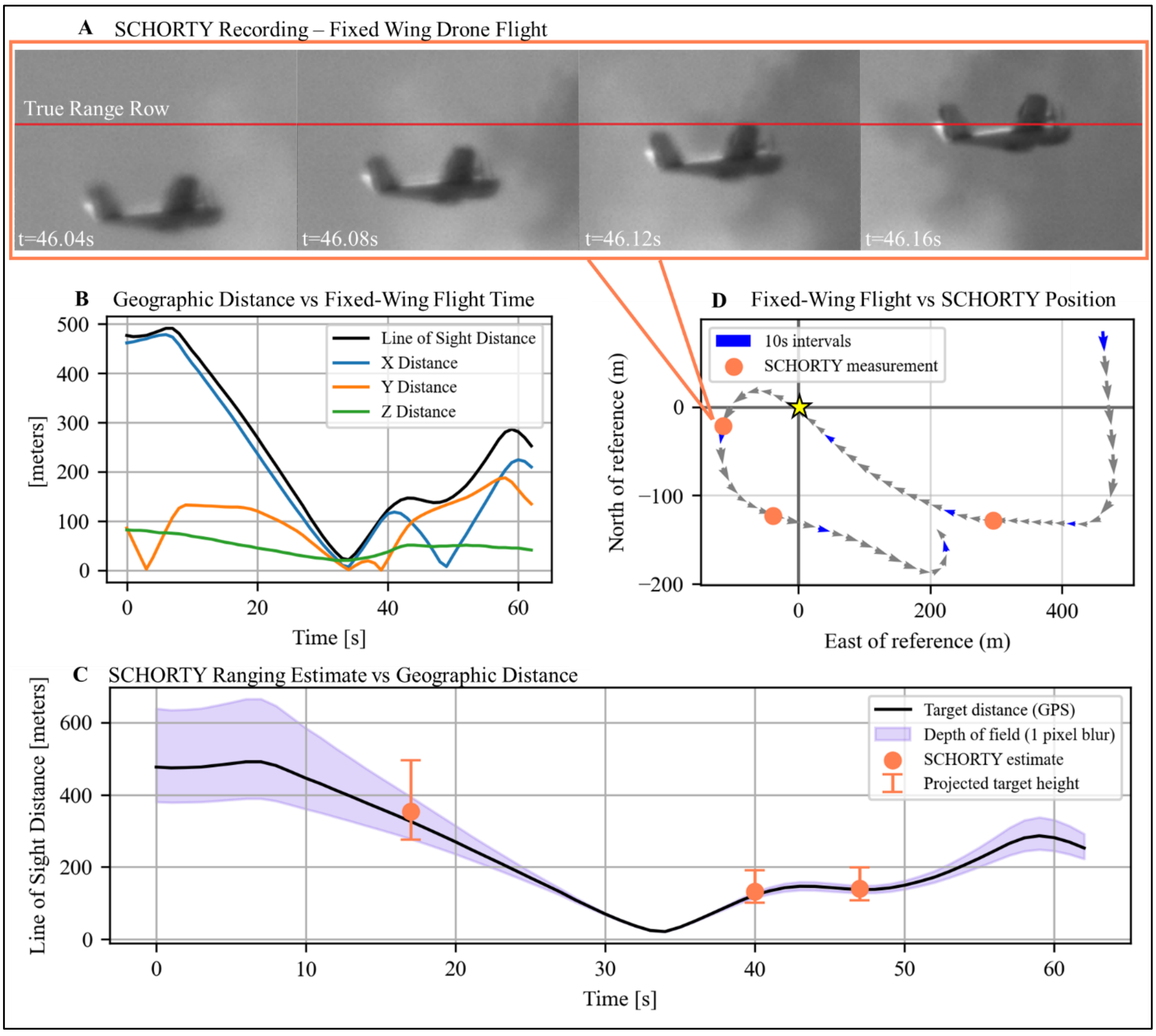

Figure 2. Fixed Wing UAV Tracking. (A) Sequential images from SCHORTY showing the UAV coming into focus as it enters the depth-mapped object plane. (B) GPS data and calculated line of sight target distance during the flight. (C) Overlay of 3 SCHORTY measurements when the UAV crossed through the object plane. (D) Calculated UAV positions using azimuth, elevation, and SCHORTY ranges overlaid on GPS flight track.

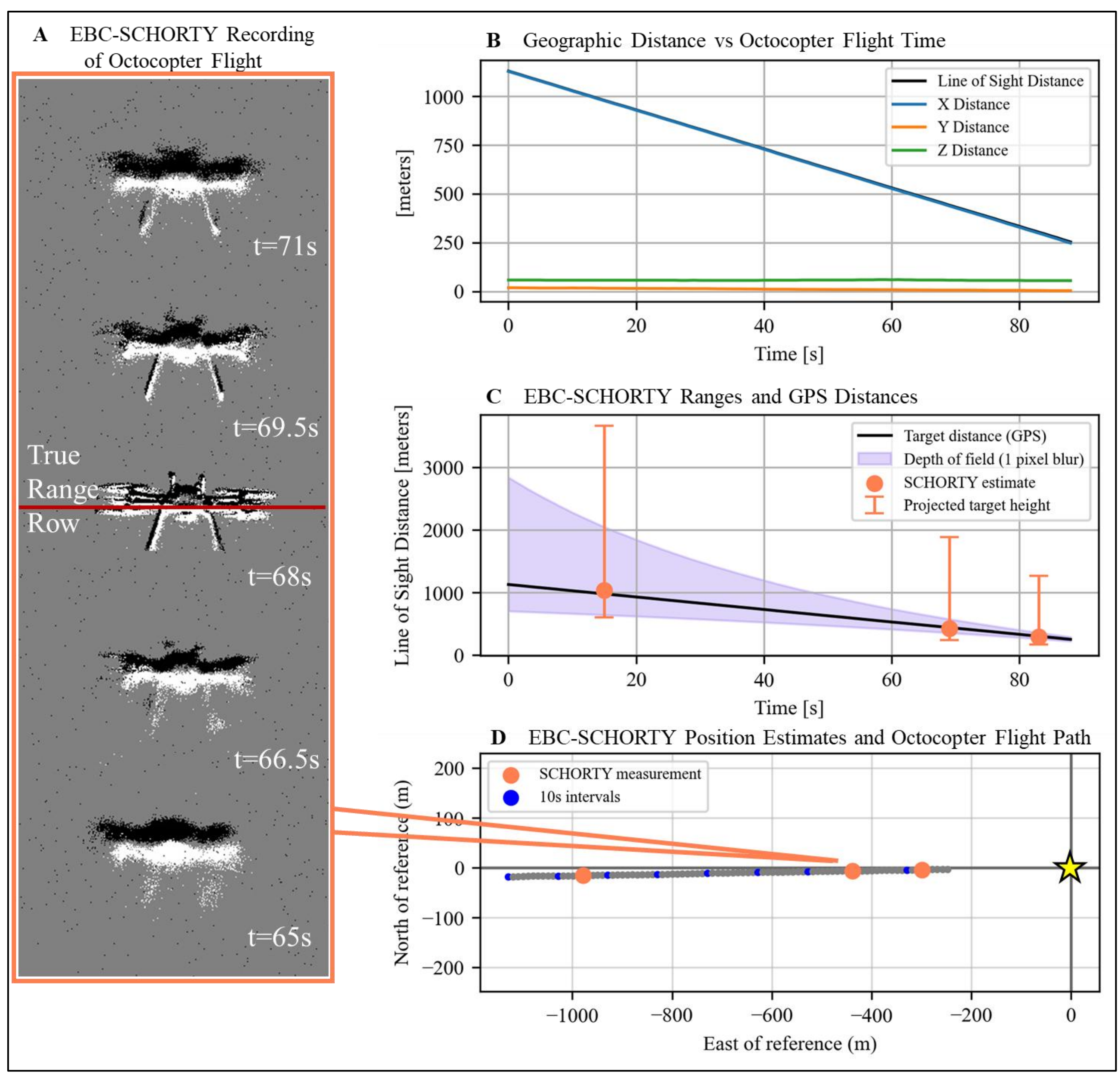

Figure 3. Octocopter UAV Tracking with EBC-SCHORTY. (A) Sequential images showing the UAV coming into focus as it enters the depth-mapped object plane. (B) GPS data and calculated line of sight target distance during the flight. Line of sight distance predominately due to X distance. (C) Overlay of 3 EBC-SCHORTY measurements when the UAV crossed through the object plane. (D) Calculated UAV positions using azimuth, elevation, and SCHORTY ranges overlaid on GPS flight track.

### 3.3 Event-based Imaging of the Octocopter with Background Suppression and Asymmetric Defocus Blurring

A hovering octocopter was recorded across an automated flight pattern with SCHORTY (see Figure 4A) and the EBC-SCHORTY (see Figure 4B). The conventional frame image captures the full scene features, including the octocopter, its rotors, and the static background environment. In contrast, the EBC inherently suppresses the static background and isolates the octocopter and rotor signature during motion. The result is a high-contrast representation of the target with no contribution from the static background, which enhances object detection in cluttered environments. Within this event-based representation, the drone was monitored as its path intersected the sharp, in-focus Scheimpflug plane and over various levels of defocus (see Figure 4C). Two observations emerge from this recording. First, as the octocopter defocuses, the event-based representation losses sharp contrast and events spread in accordance with the defocus blur in the underlying optical system. Second, the defocus exhibits an asymmetric blur size as the octocopter travels a fixed distance on either side of the Scheimpflug plane, as shown in Figure 4C.

This behavior arises from the octocopter approaching the Scheimpflug plane at an approach vector non-orthogonal to the tilt angle. Under this geometry, the projected defocus blur size exhibits an asymmetric blur size at a fixed interval in front and behind the Scheimpflug plane, as shown in Figure 4D. This property, if rigorously characterized, could be exploited to support two capabilities. First, this property may improve axial resolution by enabling precise characterization of the axial position off the focal plane. Second, by comparing the discrepancy in the blur size as a function of pixel coordinate before and after the focal plane, the angle of the drone's trajectory may be extracted, enabling a low-compute alternative to tracking in addition to range extraction. Both of these functionalities require a controlled study with targets of various approach angles across calibrated depths to further validate this interpretation.

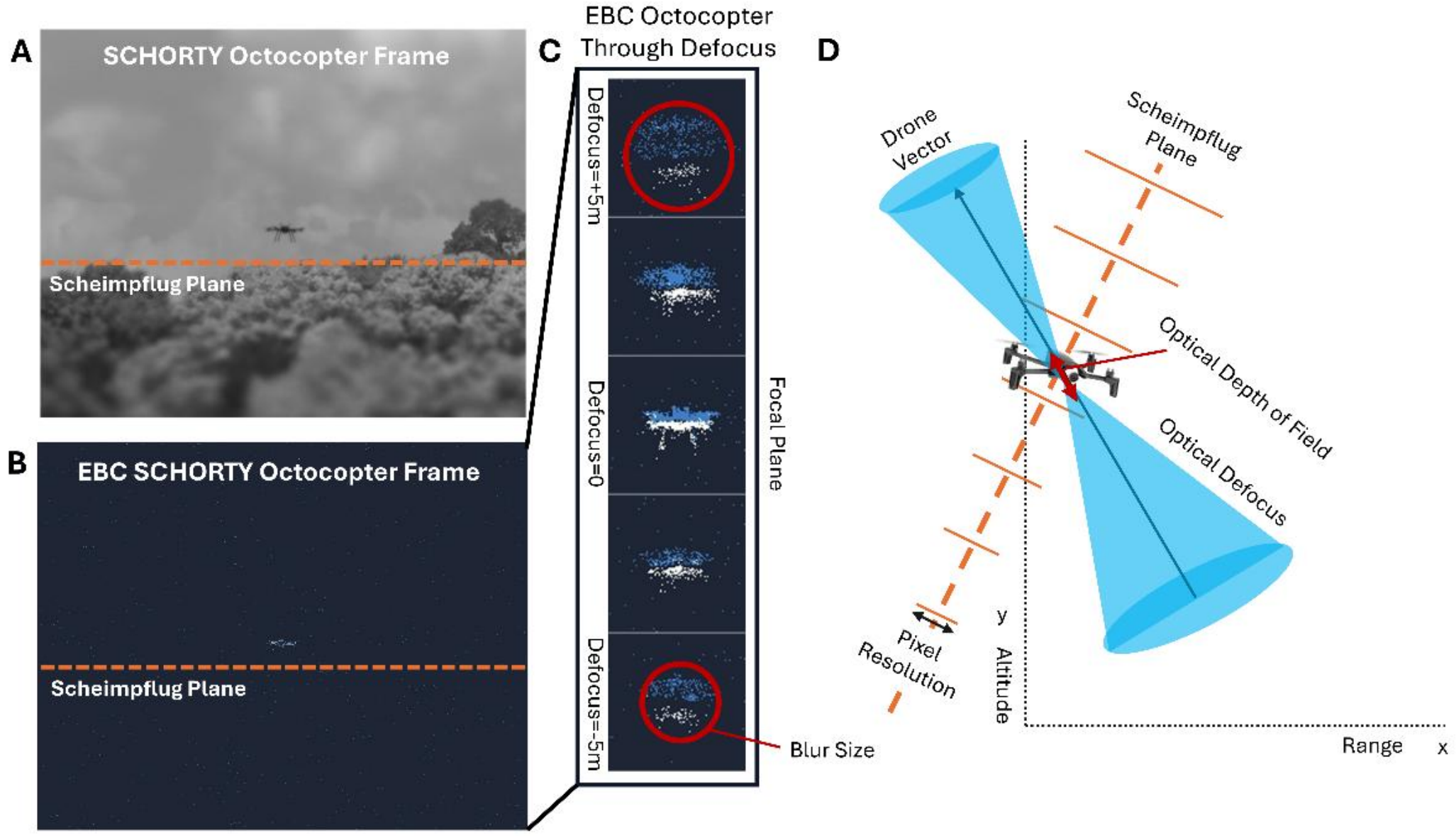

Figure 4. EBC-SCHORTY Rangefinding of the Octocopter in Flight. (A) SCHORTY and (B) EBC-SCHORTY recording of the octocopter in flight. (C) Asymmetric defocus blur recorded of drone flying through the Scheimpflug plane in event-space. (D) Depiction of how drone flight trajectory influence defocus blur across Scheimpflug plane.

## 4. DISCUSSION

This work demonstrates the application of the Scheimpflug principle to passive 3D range finding of UAVs against a natural scene. GPS coordinates are used to monitor the geographic distance between the prototype, SCHORTY, and the UAVs to establish a baseline for range finding. Range may be measured with certainty within the limits defined by the projected pixel size at a given distance, which has been demonstrated at distances up to 1.0 km. The method requires a one-time calibration to measure and fit known points within the FoV to the Scheimpflug model [14] and adjust sensor tilt and offset values accordingly to account for assembly error. SCHORTY, as a platform, enables a sensor-agnostic method for passive range finding, demonstrated by replacing the visible sensor with an EBC. In its EBC configuration, EBC-SCHORTY inherently suppresses background and accentuates motion to reduce confusers and highlight non-stationary objects. Due to the tilted imaging geometry, it was observed that defocus blurring before and after the Scheimpflug object plane introduces an asymmetry based on the UAV trajectory, which may be used to enhance localization and predict trajectory off the focal plane. However, this observation is preliminary.

When compared to other range finding strategies, SCHORTY provides a motivating trade space. The ranging relies on the direct mapping of pixels to a known range to assign distance to in-focus features, without the need for computationally intensively inverse algorithm [12]. These properties may prove advantageous when providing ranging in scenarios that benefit from low SWaP solutions, such as an all-passive alternative for mobile platforms [19] or when used in low resource environments like space-based awareness [20]. By relying on a camera-native architecture, SCHORTY can be paired with a wide range of sensors and imaging systems across modalities and wavebands to enable additional features without redesigning the underlying optical platform. Additionally, SCHORTY can be augmented with active illumination of the object plane for enhanced image contrast and LiDAR-like direct range extraction.

However, SCHORTY's range resolution is limited by the geometric projection of pixels onto the depth-mapped object plane. The resolution scales quadratically with range, such that the demonstrated system exhibits a 25 m projected pixel size at 500 m and a 100 m projected pixel size at 1 km. In addition to pixel-based limitations, the optical signal is subject to defocus when placed off the ideal object plane. Defocus introduces both ambiguity in position across the focal plane and degrades the quality of the optical signal as function of defocus distance [21]. While the EBC-SCHORTY's contrast-driven mechanism exhibited the ability to extract unique defocus profiles through the object plane, this behavior is reliant on the target trajectory and will necessitate the development of EBC deconvolution algorithms to maximize this benefit [22].

In future iterations, SCHORTY may incorporate mechanisms to aid in resolving its uncertainty when imaging off its image plane. Incorporating 2.5D optics may mitigate defocus over the pixel bin to maximize resolving objects without sacrificing conditioning at the cost of limiting ranging to the projected pixel size [23], [24]. Alternative, incorporating 3D PSF engineering techniques may enable tighter axial localization at the cost of computational cost [9]. Additionally, EBC-SCHORTY may be optimized to encode objects of expected contrast levels under different scenarios to yield a more robust ability to extract targets of interest [25]. In general, we believe addressing these challenges will continue to validate SCHORTY as a low SWaP passive ranging solution with general application across remote sensing.

## ACKNOWLEDGEMENTS

We would like to thank the Georgia Tech Research Institute Independent Research and Development (IRAD) for providing support for this project.

## REFERENCES

[1] Y. Wu *et al.*, "A Review of Intelligent Vision Enhancement Technology for Battlefield," *Wirel. Commun. Mob. Comput.*, vol. 2023, no. 1, p. 6733262, 2023, doi: 10.1155/2023/6733262.

[2] T. Ito, "Intelligence Failures in the Asymmetric War between Ukraine and Russia: A Literature Review of Ukraine's Drone Attacks on Russian Military Infrastructure," *Secur. Intell. Terror. J. SITJ*, vol. 2, no. 3, pp. 262–273, Sep. 2025, doi: 10.70710/sitj.v2i3.63.

[3] I. P. A. Kaisto, J. T. Kostamovaara, I. Moring, and R. A. Myllylae, "Laser range-finding techniques in the sensing of 3-D objects," in *Sensing and Reconstruction of Three-Dimensional Objects and Scenes*, SPIE, Jan. 1990, pp. 122–133. doi: 10.1117/12.20011.

[4] S. Se and N. Pears, "Passive 3D Imaging," in *3D Imaging, Analysis and Applications*, N. Pears, Y. Liu, and P. Bunting, Eds., London: Springer, 2012, pp. 35–94. doi: 10.1007/978-1-4471-4063-4_2.

[5] R. I. Hartley and P. Sturm, "Triangulation," *Comput. Vis. Image Underst.*, vol. 68, no. 2, pp. 146–157, Nov. 1997, doi: 10.1006/cviu.1997.0547.

[6] R. A. Raynor, "Range Finding with a Plenoptic Camera," Art. no. AFITENP14M29, Mar. 2014, Accessed: Apr. 17, 2026. [Online]. Available: https://apps.dtic.mil/sti/html/tr/ADA599366/

[7] M. R. Rai and J. Rosen, "Depth-of-field engineering in coded aperture imaging," *Opt. Express*, vol. 29, no. 2, pp. 1634–1648, Jan. 2021, doi: 10.1364/OE.412744.

[8] Y. Zong, "Coded aperture compressive snapshot imaging for moving objects location," in *Fourth International Conference on Image Processing and Intelligent Control (IPIC 2024)*, SPIE, Aug. 2024, pp. 50–56. doi: 10.1117/12.3038535.

[9] S. Fu *et al.*, “Deformable mirror based optimal PSF engineering for 3D super-resolution imaging,” *Opt. Lett.*, vol. 47, no. 12, pp. 3031–3034, Jun. 2022, doi: 10.1364/OL.460949.
[10] K. Khare, M. Butola, and S. Rajora, “PSF Engineering,” in *Fourier Optics and Computational Imaging*, K. Khare, M. Butola, and S. Rajora, Eds., Cham: Springer International Publishing, 2023, pp. 249–260. doi: 10.1007/978-3-031-18353-9_17.
[11] J. Zhang, N. Pégard, J. Zhong, H. Adesnik, and L. Waller, “3D computer-generated holography by non-convex optimization,” *Optica*, vol. 4, no. 10, pp. 1306–1313, Oct. 2017, doi: 10.1364/OPTICA.4.001306.
[12] K. Yanny, K. Monakhova, R. W. Shuai, and L. Waller, “Deep learning for fast spatially varying deconvolution,” *Optica*, vol. 9, no. 1, pp. 96–99, Jan. 2022, doi: 10.1364/OPTICA.442438.
[13] M. Kojima, A. Wegener, and O. Hockwin, “Imaging Characteristics of Three Cameras Using the Scheimpflug Principle,” *Ophthalmic Res.*, vol. 22, no. Suppl. 1, pp. 29–35, Dec. 2009, doi: 10.1159/000267061.
[14] N. Meraz *et al.*, “Scheimpflug cameras for range-resolved observations of the atmospheric effects on laser propagation,” in *Laser Radar Technology and Applications XXX*, SPIE, May 2025, pp. 41–60. doi: 10.1117/12.3054806.
[15] G. Gallego *et al.*, “Event-Based Vision: A Survey,” *IEEE Trans. Pattern Anal. Mach. Intell.*, vol. 44, no. 1, pp. 154–180, Jan. 2022, doi: 10.1109/TPAMI.2020.3008413.
[16] T. Kim, H. Cho, and K.-J. Yoon, “Frequency-aware Event-based Video Deblurring for Real-World Motion Blur,” presented at the Proceedings of the IEEE/CVF Conference on Computer Vision and Pattern Recognition, 2024, pp. 24966–24976. Accessed: Apr. 22, 2026. [Online]. Available: https://openaccess.thecvf.com/content/CVPR2024/html/Kim_Frequency-aware_Event-based_Video_Deblurring_for_Real-World_Motion_Blur_CVPR_2024_paper.html
[17] N. Boehrer, R. P. J. Nieuwenhuizen, and J. Dijk, “Turbulence mitigation in imagery including moving objects from a static event camera,” *Opt. Eng.*, vol. 60, no. 5, p. 053101, May 2021, doi: 10.1117/1.OE.60.5.053101.
[18] I. Akanbi and M. Ayomoh, “Event-Based Vision Application on Autonomous Unmanned Aerial Vehicle: A Systematic Review of Prospects and Challenges,” *Sensors*, vol. 26, no. 1, Dec. 2025, doi: 10.3390/s26010081.
[19] J. R. Kellner *et al.*, “New Opportunities for Forest Remote Sensing Through Ultra-High-Density Drone Lidar,” *Surv. Geophys.*, vol. 40, no. 4, pp. 959–977, Jul. 2019, doi: 10.1007/s10712-019-09529-9.
[20] C. Toth and G. Jóźków, “Remote sensing platforms and sensors: A survey,” *ISPRS J. Photogramm. Remote Sens.*, vol. 115, pp. 22–36, May 2016, doi: 10.1016/j.isprsjprs.2015.10.004.
[21] Y. Xiong and S. A. Shafer, “Depth from focusing and defocusing,” in *Proceedings of IEEE Conference on Computer Vision and Pattern Recognition*, Jun. 1993, pp. 68–73. doi: 10.1109/CVPR.1993.340977.
[22] C. Haoyu, T. Minggui, S. Boxin, W. YIzhou, and H. Tiejun, “Learning to Deblur and Generate High Frame Rate Video with an Event Camera,” Mar. 20, 2020, *arXiv*: arXiv:2003.00847. doi: 10.48550/arXiv.2003.00847.
[23] F. Aguet, D. Van De Ville, and M. Unser, “Model-Based 2.5-D Deconvolution for Extended Depth of Field in Brightfield Microscopy,” *IEEE Trans. Image Process.*, vol. 17, no. 7, pp. 1144–1153, Jul. 2008, doi: 10.1109/TIP.2008.924393.
[24] J. Greene *et al.*, “Pupil engineering for extended depth-of-field imaging in a fluorescence miniscope,” *Neurophotonics*, vol. 10, no. 4, p. 044302, May 2023, doi: 10.1117/1.NPh.10.4.044302.
[25] J. L. Greene, A. Kar, I. Galindo, E. Quiles, E. Chen, and M. Anderson, “A PyTorch-enabled tool for synthetic event camera data generation and algorithm development,” in *Synthetic Data for Artificial Intelligence and Machine Learning: Tools, Techniques, and Applications III*, SPIE, May 2025, pp. 123–143. doi: 10.1117/12.3053238.